\documentclass[runningheads]{llncs}
\usepackage[T1]{fontenc}
\usepackage{graphicx}
\usepackage{bbm}
\usepackage{multirow}
\usepackage{booktabs}
\usepackage{wrapfig}
\usepackage{amssymb}
\usepackage{url}
\usepackage{bbding}
\usepackage{float} 
\usepackage{threeparttable}

\newcommand{\etal}{\textit{ et al}. }
\newcommand{\ie}{\textit{i}.\textit{e}., }

\begin{document}
\title{Cross-Fundus Transformer for Multi-modal Diabetic Retinopathy Grading with Cataract}
\titlerunning{CFT for Multi-modal DR Grading with Cataract}

\author{Anonymous}
\institute{Anonymous}
\author{Fan Xiao\inst{1} \and
Junlin Hou\inst{2} \and
Ruiwei Zhao\inst{1} \and 
Rui Feng\inst{1,2*} \and
Haidong Zou\inst{3} \and
Lina Lu\inst{3} \and
Yi Xu\inst{3} \and
Juzhao Zhang\inst{4}}
\authorrunning{Xiao et al.}
%
\institute{Academy for Engineering and Technology, Fudan University, Shanghai, China \email{21210860085@m.fudan.edu.cn, \{rwzhao, fengrui\}@fudan.edu.cn} \and
School of Computer Science, Fudan University, Shanghai, China \email{18110240004@fudan.edu.cn}\\
\and
Shanghai Eye Diseases Prevention \& Treatment Center, Shanghai, China \and
Shanghai Jiao Tong University School of Medicine, Shanghai, China}

\maketitle              
\begin{abstract}
Diabetic retinopathy (DR) is a leading cause of blindness worldwide and a common complication of diabetes. 
As two different imaging tools for DR grading, color fundus photography (CFP) and infrared fundus photography (IFP) are highly-correlated and complementary in clinical applications.  
To the best of our knowledge, this is the first study that explores a novel multi-modal deep learning framework to fuse the information from CFP and IFP towards more accurate DR grading.
Specifically, we construct a dual-stream architecture Cross-Fundus Transformer (CFT) to fuse the ViT-based features of two fundus image modalities. In particular, a meticulously engineered Cross-Fundus Attention (CFA) module is introduced to capture the correspondence between CFP and IFP images. Moreover, we adopt both the single-modality and multi-modality supervisions to maximize the overall performance for DR grading.
Extensive experiments on a clinical dataset consisting of 1,713 pairs of multi-modal fundus images demonstrate the superiority of our proposed method.
Our code will be released for public access.

\keywords{Multi-modal Fundus Image \and Diabetic Retinopathy \and Cross-Fundus Transformer.}
\end{abstract}
\section{Introduction}

\begin{figure}[t!]
    \includegraphics[width=0.95\textwidth]{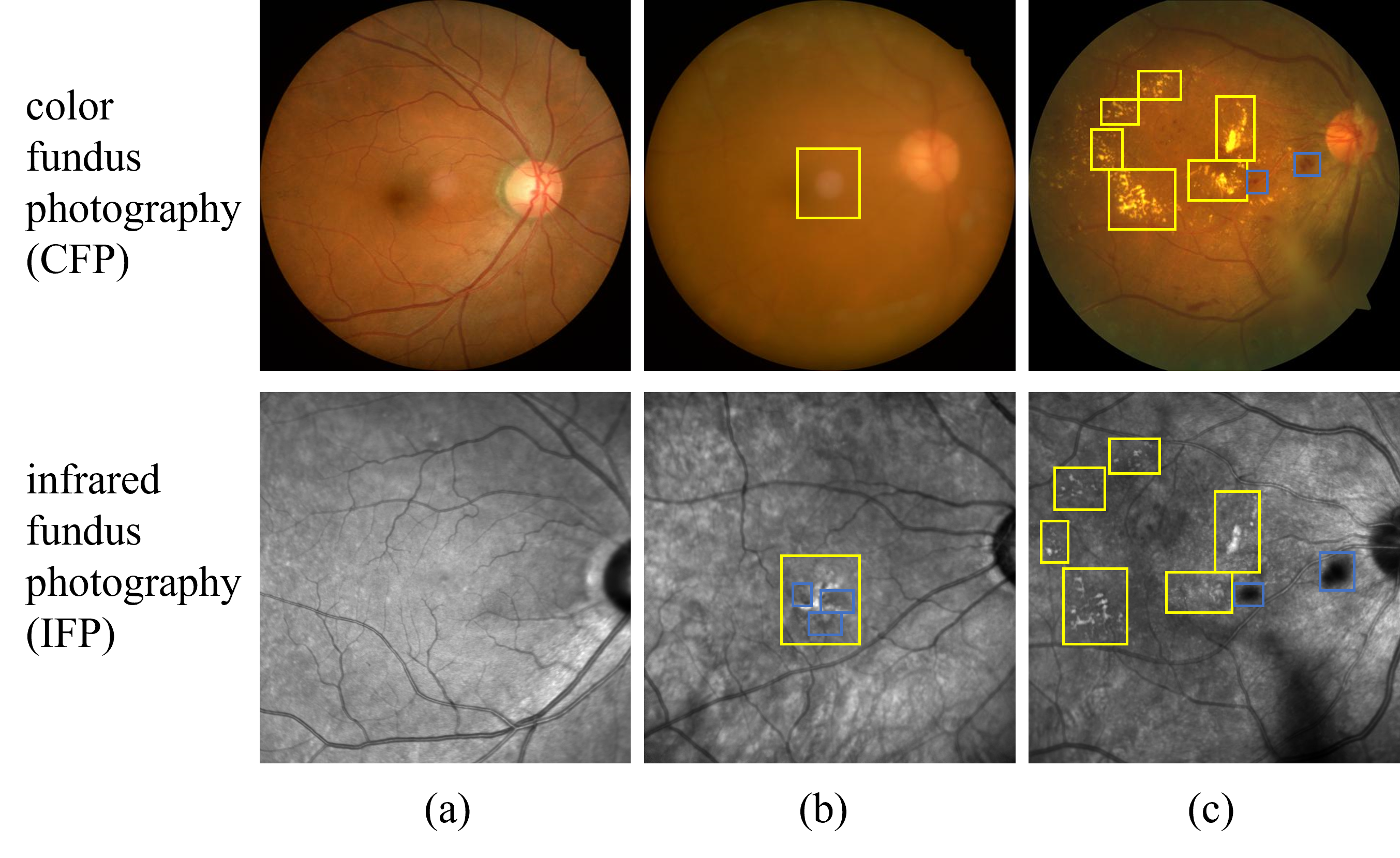}
    \caption{Examples of CFP and IFP image. (a) Both clear images; (b) The same lesion is unclear in CFP while clear in IFP; (c) Comparison of different DR lesions in CFP and IFP.}
    \label{intro}     
\end{figure}

Diabetic retinopathy (DR) is one of the chronic complications of diabetes mellitus and the leading cause of avoidable blindness \cite{sayin2015ocular}. 
According to the International Clinical Diabetic Retinopathy Scale \cite{wilkinson2003proposed}, the severity of DR is graded into five grades, \ie no DR, mild Nonproliferative DR (NPDR), moderate NPDR, severe NPDR, and proliferative DR (PDR).
In clinical practice, color fundus photography (CFP) is commonly used for DR screening.
However, patients with cataract have serious lens opacity, which obviously affects the imaging effect of CFP. Instead, infrared fundus photography (IFP) can obtain clear retinal images due to the good penetration of infrared light to turbidity media
Thus, IFP can complement other imaging techniques in monitoring and assessing treatment response in DR patients \cite{xue2022deep,sukkarieh2022role,roh2021infrared}.
Fig. \ref{intro} shows some examples of CFP and IFP images. 
Fig. \ref{intro}a illustrates that CPF and IFP of healthy people have clear fundus structure which exhibits a significant correlation in terms of image resolution and utility \cite{ajaz2019relation}.
On one hand, Fig. \ref{intro}b captured from a cataract patient shows that CFP is obscured by a gray fog, while IFP remains clear enough to discern the fundus structure.
Particularly, useful features for DR diagnosis such as hemorrhages (blue box) and exudate (yellow box) can be more easily identified in IFP.
On the other hand, as illustrated in Fig. \ref{intro}c that there are appearance differences of lesions between the two modalities. 
Specifically, CFP provides better color information that aids doctors in distinguishing lesions, which are often overlooked in IFP \cite{sukkarieh2022role}.
This dichotomy underscores the potential of leveraging the complementary strengths of both  modalities to enhance DR grading accuracy.

\begin{figure*}[t]
\centering
\includegraphics[width=\textwidth]{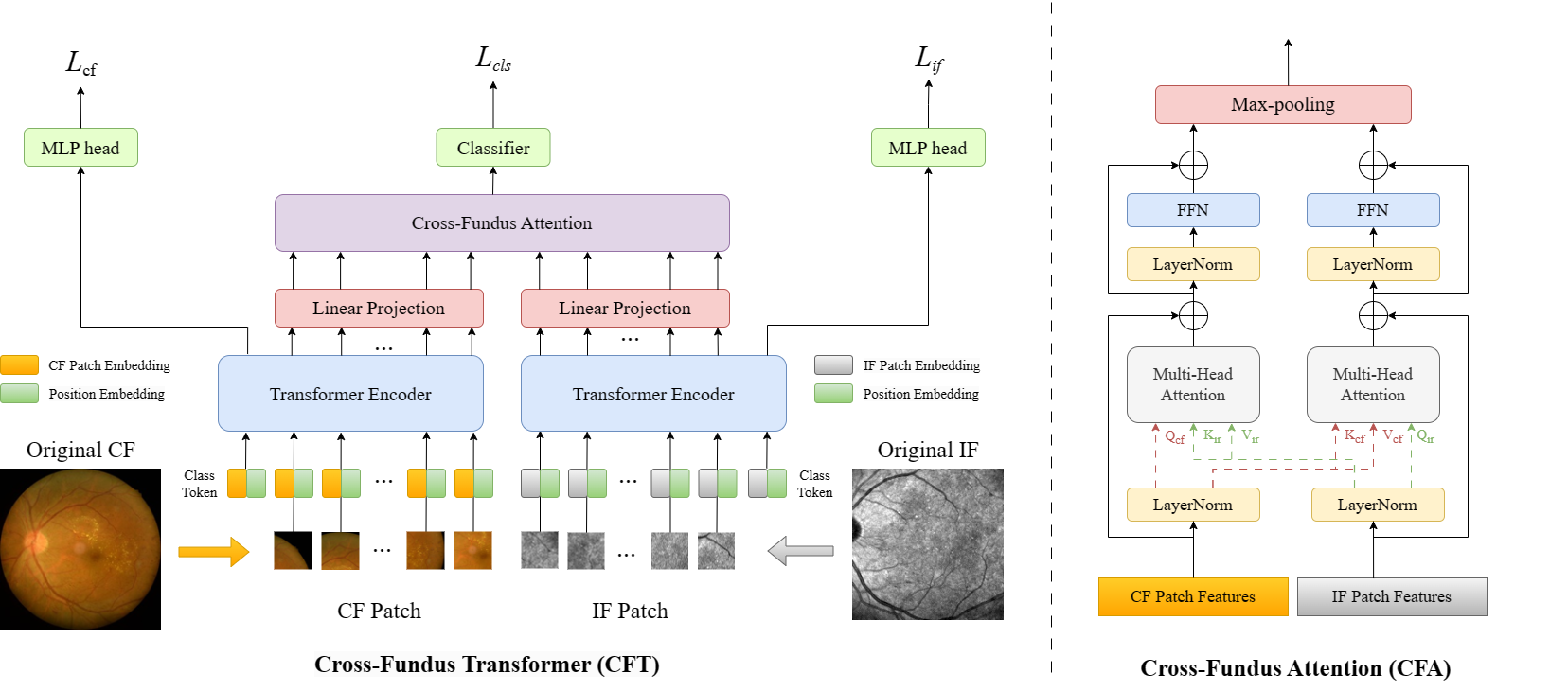} 
\caption{The architecture of our Cross-Fundus Transformer (CFT) for DR grading. It consists Transformer encoders, linear projection layers,  Cross-Fundus Attention (CFA) module, classifier layer and MLP heads.} 
\label{network}
\end{figure*}
In recent years, there has been an increase in the use of multi-modal fundus images for retinal disease diagnosis. 
Most existing works focus on the fusion of CFP and optical coherence tomography (OCT). 
For example, Wang\etal\cite{10388423} proposed  a multi-modal learning network, termed GeCoM-Net, which encodes the geometric correspondences between OCT slices and  their associated CFP regions. This approach advances  the diagnosis of retinal diseases such as diabetic macular edema, impaired visual acuity and glaucoma, demonstrating enhanced diagnostic efficacy.
Other works consider the differences between modalities and design tailored fusion strategies \cite{wang2019two,li2021multi,he2021multi}.
However, multi-modal interaction of CFP and IFP for retinal disease diagnosis is still largely unexplored.

In this study, we propose a novel dual-stream network architecture, named Cross-Fundus Transformer (CFT), based on a vision transformer (ViT) \cite{dosovitskiy2020image} to diagnose DR using both CFP and IFP images. 
Our approach involves generating CFP and IFP tokens using ViT, which are then fused by a Cross-Fundus Attention (CFA) module to obtain a unified representation for DR grading. 
We particularly construct a clinical dataset, consisted of 1,713 pairs of CFP and IFP images, to evaluate the effectiveness of our algorithm.
Our work represents the first attempt to automatically diagnose DR using both CFP and IFP medical images, and our experimental results demonstrate state-of-the-art performance.

\section{Methodology}
Based on multi-modal fundus imaging comprised of CFP and IFP, we propose a novel dual-stream network CFT for DR grading.
In particular, we design a CFA module that fully leverages the feature representations extracted by individual patches of CFP and IFP images. 
The fused representations along with the overall representation obtained by the class token is then classified to obtain the DR grading result. 

\subsection{Network architecture}

As shown in Fig. \ref{network}, the input CFP image $\mathbf{I}_{cf}\in \mathbbm{R}^{H\times W\times C}$ and IFP image $\mathbf{I}_{if}\in \mathbbm{R}^{H\times W\times C}$are first divided into patches with size $p_{cf}$ and $p_{if}$.
Then the number of patches $N$ can be calculated by $N = (H\times W)/ p^2$. 
These patches of the two fundus modalities are flattened into one-dimensional vectors and then passed through the embedding composed of linear layers.
The obtained patch embedding concatenates the position embedding which is a learned one-dimensional vector in order to get position information.
Because of the overall image information to be considered, a learnable class token is the same as patch embedding concatenating the position embedding and serving as the first one in the embedding sequence.
The embedding sequence is fed into ViT encoder which contains repeated Transformer blocks. 
Specifically, each Transformer block consists of multi-head attention (MHA) and feed forward network (FFN), which are connected by residual operations and layer normalization.

The input CFP image $\mathbf{I}_{cf}$ and IFP image $\mathbf{I}_{if}$ through the above process are encoded to the feature representations.
Then the feature representations of the class token are fed into a multi-layer perception for single modality DR grading.
The other feature representations corresponding to patches are fed into CFA module to obtain cross-modality information.
Then the fused patch feature representations are fed into a classifier consisting of a layer normalization and a linear layer for multi-modality DR grading.

\subsection{Cross-Fundus Attention Module}
In order to integrate the representation of the patch embedding between the two modalities, we design a CFA module to obtain fusion information. 
All of the feature representations of the patch embedding $\mathbf{F}_{cf}\in \mathbbm{R}^{N_{cf}\times C}$ and $\mathbf{F}_{if}\in \mathbbm{R}^{N_{if}\times C}$are fed into linear projection to reduce feature dimension to $L$, where $N$ represents the number of patches and $C$ represents the dimension. 
The linear projection block consists of a linear layer, a layer normalization, and a ReLU activation function.
The new feature representations of the patch embedding $\mathbf{F'}_{cf}\in \mathbbm{R}^{N_{cf}\times L}$ and $\mathbf{F'}_{if}\in \mathbbm{R}^{N_{if}\times L}$ go through a cross attention block, as shown in Fig. \ref{network}.
We split the queries, keys, and values between the two modal representations of the patch embeddings.
Specifically, we denote $\mathbf{F'}_{cf}$ as queries, $\mathbf{F'}_{if}$ as keys and values in CF-cross attention and $\mathbf{F'}_{if}$ as queries, $\mathbf{F'}_{cf}$ as keys and values in IF-cross attention:
\begin{equation}
   Q_{cf}=\mathbf{F'}_{cf}W^Q, K_{cf}=\mathbf{F'}_{if}W^K, V_{cf}=\mathbf{F'}_{if}W^V, 
\end{equation}
\begin{equation}
    Q_{if}=\mathbf{F'}_{if}W^Q, K_{if}=\mathbf{F'}_{cf}W^K, V_{if}=\mathbf{F'}_{cf}W^V, 
\end{equation}
 where $W^Q,W^K,W^V\in \mathbbm{R}^{L \times{d}}$ are linear projections.
Then, due to the multi-head mechanism we split the converted $Q, K, V\in \mathbbm{R}^{N\times d}$ into $\{Q_n\}_{n=1}^N$, $\{K_n\}_{n=1}^N$, $\{V_n\}_{n=1}^N$ where $N$ represents the number of heads.
So we can calculate the attention weight of each head by 
\begin{equation}
    A_n = softmax(\frac{Q_nK_n^T}{\sqrt{d/N}}), 
\end{equation}
where $\sqrt{d/N}$ represents a scaling factor.
The output of head $H_n \in \mathbbm{R}^{N\times d/N}$ is calculated by matrix multiplication of the attention weight $A_n$ and values $Vn$, \ie $H_n=A_nV_n$.
Finally, we concatenate the output of each single head $\{H_n\}_{n=1}^N$ along the channel dimension to obtain the multi-head output $H \in \mathbbm{R}^{N\times d}$ through a linear projection matrix $W^O \in \mathbbm{R}^{{d}\times L}$, 
\begin{equation}
    H=concat[H_1;H_2;...;H_N]W^O.
\end{equation}
The multi-head output is then fed into FFN with two linear layers and a GELU activation function.
Until now, the two single feature representations of the patch embedding between CFP and IFP are obtained and fed into a max-pooling layer for final fused representations.

\subsection{Training and Inference}
To speed up the training process, we utilize a ViT model pretrained on a large fundus image dataset \cite{yu2021mil}, and finetune it on our target data.
We adopt the cross entropy loss function to train the single modality `MLP head' and fusion classifier:
\begin{equation}
    L(y,\hat{y}) = -\sum\limits_{i=1}^{k}y_ilog(\hat{y}_i),
\end{equation}
where $y$ represents the ground-truth label and $\hat{y}$ represents the prediction label.
The total loss is comprised of two single-modal losses $L_{cf},L_{if}$ and a multi-modal loss $L_{cls}$: 
\begin{equation}
    L_{total} = \lambda L_{cf}+ (1-\lambda)L_{if} + L_{cls},
\end{equation}
where $\lambda$ is the hyper-parameter that balances the loss of the two modalities.
In the inference stage, the output of two heads is averaged and added to the output of the classifier to get the final result.

\section{Experiments}
\subsection{Clinical Dataset}
Due to the lack of publicly available CFP and IFP image pair datasets, we evaluate our method on a clinically acquired dataset collected from Department of Ophthalmology of the 
anonymous Hospital from Jan. 2020 to March. 2022. 
The dataset named CFP-IFP DR (CIDR) comprises  paired CFP and IFP images from 1,713 eyes of 616 patients in different periods, including instances of patients afflicted with cataract.
CFP images were captured by a fundus camera with a resolution over 2,000$\times$2,000, and IFP images were obtained from a retinal health assessment device with a resolution of 768$\times$768.
Each pair of images was annotated by an experienced ophthalmologist.
Eventually, we obtain the labeled CIDR with 714 eyes showing no DR, 123 eyes showing mild NPDR, 249 eyes showing moderate NPDR, 267 eyes showing severe NPDR, and 360 eyes showing PDR.
We randomly split CIDR into 80\% for training and 20\% for validation according to category proportions.

\subsection{Implementation details}
All networks are optimized using Adam optimizer with a weight decay of 1e-5 for 100 epochs. The initial learning rate is set to 1e-4 with cosine annealing. Both CFP and IFP images are resized to 512$\times$512, and data augmentations include flipping, random cropping, color jittering, and affine transformations.
We use the Quadratic Weighted Kappa \cite{graham2015kaggle}, Accuracy, and Macro-F1 score as overall comparison metrics.

\begin{table}[t]
\centering
\caption{Comparison to state-of-the-art methods. (UNIT: \%)}
\label{comparision}
\begin{tabular}{clccc}
\toprule
~~Modal~~ &  Method & ~Kappa~ & ~~Acc~~ & ~~~~F1~~~~ \\
\midrule
\multirow{4}{*}{Single} & CFP & 79.06 & 64.43 &52.39\\
& IFP & 79.19 & 68.22 &58.57\\
& CFP-self attention & 80.03 & 65.31 &55.44\\
& IFP-self attention & 79.83 & 66.18 &54.73\\
\midrule
\multirow{7}{*}{Multi}&voting-max \cite{goatman2011assessment} &81.45&67.93&55.89\\
&voting-average \cite{wu2019deep} &82.19&70.26&59.90\\
&feature-maxpool \cite{hou2022cross} &81.85&67.64&57.36\\
&feature-meanpool \cite{hashir2020quantifying} &81.50&67.35&58.22\\
&feature-concatenate \cite{wang2019two} &81.97&66.47&56.74\\
\cmidrule{2-5}
&CFP-cross attention&82.58&67.93&56.15\\
&IFP-cross attention&82.86&69.68&60.55\\
&CFP-IFP-cross attention~ &\textbf{84.44}&\textbf{73.47}&\textbf{65.51}\\
\bottomrule
\end{tabular}
\end{table}


\subsection{Comparison to other state-of-the-art methods}
We conduct experiments on our CIDR dataset. 
For single-modal methods, we train ViT and ViT with self attention at patch features of last layer using CFP and IFP images, respectively.
For multi-modal methods, we compare our model with commonly-used fusion strategies, including voting \cite{goatman2011assessment,wu2019deep} and feature fusion \cite{hou2022cross,hashir2020quantifying,rubin2018large}.
We adopt the ViT-Small structure as our backbone.
As can be observed from the single-modal section in Table \ref{comparision}, 
compared with ViT models, two ViT-self attention models increase the kappa scores by 0.97\% and 0.64\% on CFP and IFP, respectively.
Furthermore, the group of multi-modal approaches show significant performance improvement, compared to all single-modal approaches.
Among the multi-modal fusion methods, our proposed CFP-IFP-cross attention model achieves the best performance with a Kappa score of 84.44\%, an Accuracy score of 73.47\% and a F1 score of 65.51\%.
Besides, dual-stream cross attention architecture is better than single-stream cross attention architecture with an increase of 1.86\% and 1.56\% on Kappa,  respectively.
This suggests that the multi-modal fundus images have complementary features, which can improve the DR grading accuracy.

\subsection{Ablation Study}

Furthermore, we conduct an ablation study to evaluate the efficacy of linear projection, loss function, and fusion methods.
\begin{table}[t]
\centering
\caption{Ablation study of loss function, linear projection (LP), and fusion methods. (UNIT: \%)}\label{ablation}
\begin{tabular}{c|cccc|c|c|c}
\hline
~ID~ & ~$L_{cf}$~ & ~$L_{if}$~ & ~LP~ & ~Fusion~ & ~Kappa~ & ~~~Acc~~~ & ~~~F1~~~ \\
\hline
1&&&&max&81.33&67.64&58.00\\
2&&&\Checkmark&max&81.56&66.76&56.12\\
3&\Checkmark&&\Checkmark&max&82.22&67.06&57.67\\
4&&\Checkmark&\Checkmark&max&82.54&67.64&57.93\\
5&\Checkmark&\Checkmark&\Checkmark&mean&83.08&69.97&60.63\\
6&\Checkmark&\Checkmark&\Checkmark&concat&83.69&70.85&63.66\\
7&\Checkmark&\Checkmark&\Checkmark&max&\textbf{84.44}&\textbf{73.47}&\textbf{65.51}\\
\hline
\end{tabular}
\end{table}
The first two rows of Table \ref{ablation} demonstrate the effectiveness of linear projection with an increase of 0.23\% on Kappa.
The loss functions $L_{if}$ and $L_{cf}$ bring 0.86\% and 0.54\% improvements of Kappa respectively in lines 3 to 7 and 4 to 7 of the Table \ref{ablation}.
These findings show that the linear projection structure with $L_{if}$ and $L_{cf}$ function is effective in enhancing the performance of the DR grading.
Furthermore, we compare the maxpool fusion operation in our CFA with two common feature fusion methods, i.e., meanpool and concatenate. 
The Kappa score of the maxpool method is higher than the other two feature fusion methods by 1.36\% and 0.75\% in lines 7, 5 and 6.
This indicates that maxpool fusion method can better integrate features between the two modalities.

\begin{figure}[t!]
   \centering
    \includegraphics[width=0.9\textwidth]{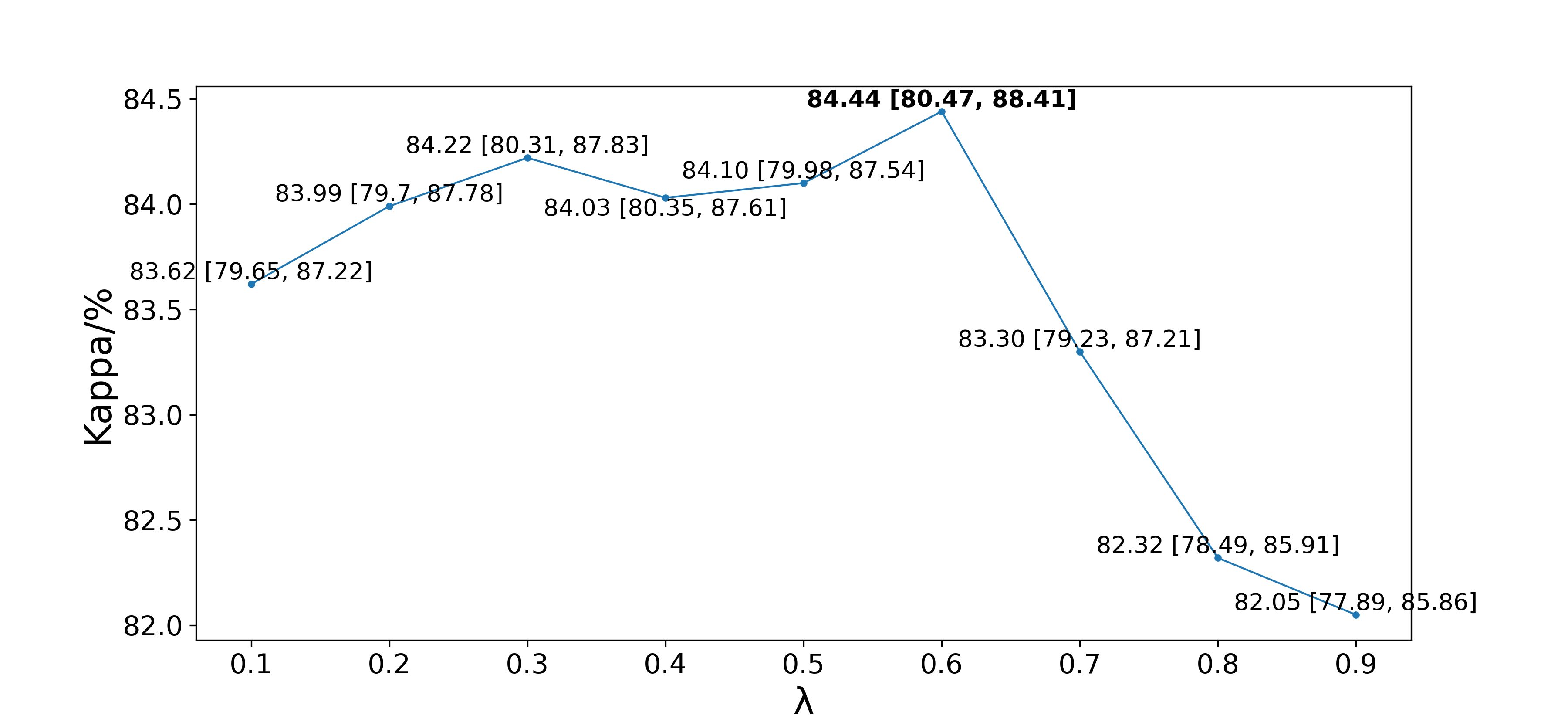}
    \caption{Results of different loss function weights $\lambda$. The numbers in square brackets indicate the 95\% confidence interval}
    \label{parameter}
\end{figure}

We also investigated the influence of different sets of loss weights.
As shown in Fig. \ref{parameter}, loss function weight $\lambda$ get the best performance at 0.6, with a kappa score of 84.44\%.
Excessively high or low weights would significantly affect the performance and generalizability of the network, ultimately hindering its ability to achieve optimal results. 
More supervision information on CFP is beneficial, but lack of supervision information on IFP image will also lead to poor effect.

\subsection{Visualization of Attention Features}
\begin{figure}[t]
\centering
\includegraphics[width=\textwidth]{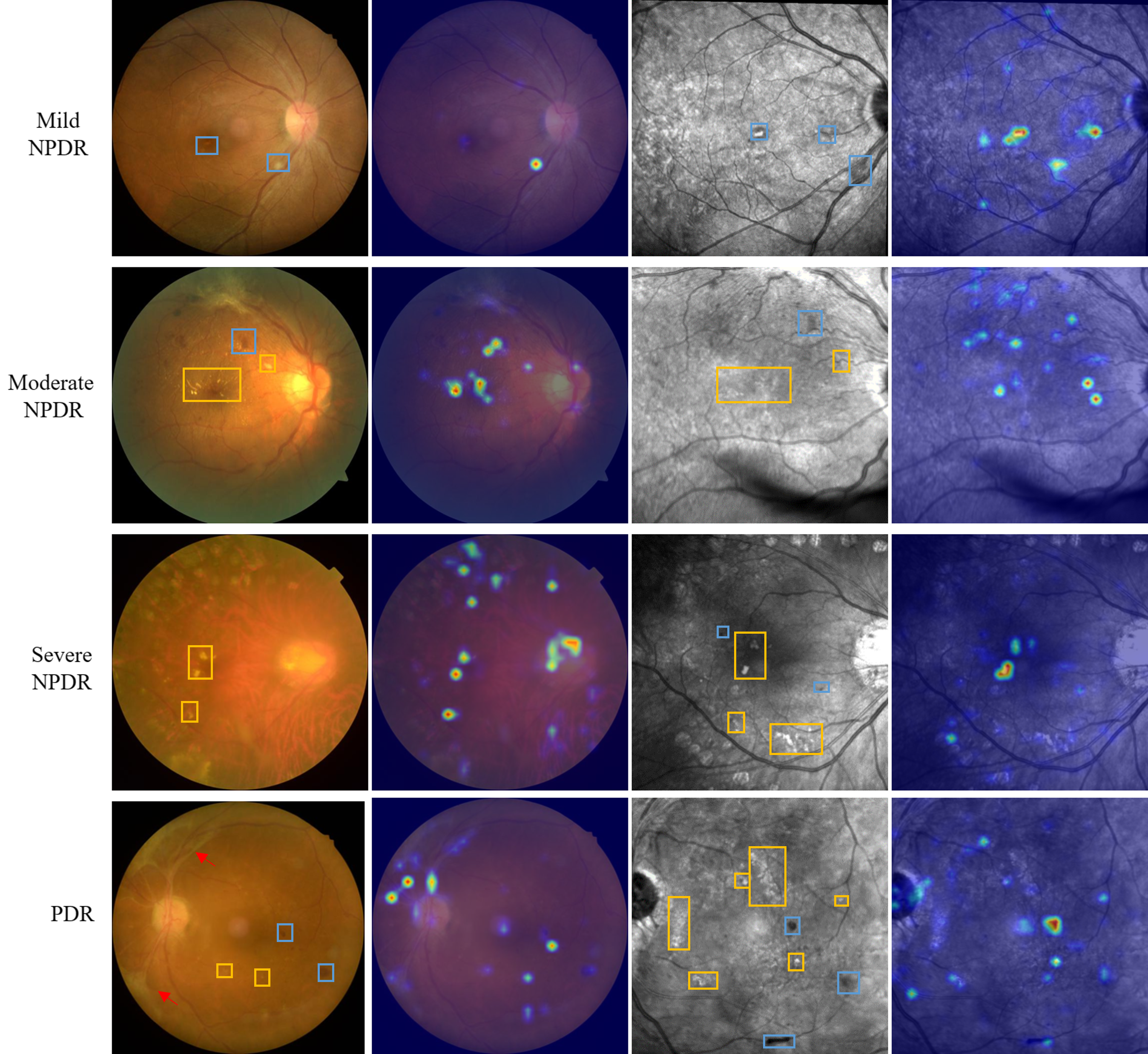} 
\caption{Visualization of CFP and IFP for DR grading via Attention Rollout \cite{abnar2020quantifying}.} 
\label{rollout}
\end{figure}

Finally, we present visualization results from the `MLP head' by Attention Rollout \cite{abnar2020quantifying} in Fig. \ref{rollout}.
There are different but complementary highlighted lesion regions in CFP and IFP.
Lesions like retinal detachment (red arrow) is easily distinguished in CFP at PDR stage, while other lesions such as exudate (yellow box) and hemorrhages (blue box) are more easily distinguished in IFP.
Therefore, fusing this complementary information enhances DR grading accuracy.

\section{Conclusion}
In this work, we constructed a novel multi-modal framework CFT for DR grading on CFP and IFP images.
We used ViT to extract features from patches in each modality, and integrated features across modalities using a designed CFA module. 
Experiments on a clinical dataset showed that our model achieved state-of-the-art results compared to existing methods.
Overall, our study presents a promising step towards multi-modal fundus image diagnosis and has the potential to contribute to the advancement of medical image analysis.
Overall, our study presents a promising step towards multi-modal fundus image diagnosis and has the potential to contribute to the advancement of medical image analysis.


%
%
%
%
\bibliographystyle{splncs04}
\bibliography{mybib}

\begin{thebibliography}{10}
\providecommand{\url}[1]{\texttt{#1}}
\providecommand{\urlprefix}{URL }
\providecommand{\doi}[1]{https://doi.org/#1}

\bibitem{abnar2020quantifying}
Abnar, S., Zuidema, W.: Quantifying attention flow in transformers. arXiv preprint arXiv:2005.00928  (2020)

\bibitem{ajaz2019relation}
Ajaz, A., Aliahmad, B., Sarossy, M., Kumar, D.K.: Relation between retinal vasculature and retinal thickness in macular edema. In: 2019 41st Annual International Conference of the IEEE Engineering in Medicine and Biology Society (EMBC). pp. 5593--5596. IEEE (2019)

\bibitem{dosovitskiy2020image}
Dosovitskiy, A., Beyer, L., Kolesnikov, A., Weissenborn, D., Zhai, X., Unterthiner, T., Dehghani, M., Minderer, M., Heigold, G., Gelly, S., et~al.: An image is worth 16x16 words: Transformers for image recognition at scale. arXiv preprint arXiv:2010.11929  (2020)

\bibitem{goatman2011assessment}
Goatman, K., Charnley, A., Webster, L., Nussey, S.: Assessment of automated disease detection in diabetic retinopathy screening using two-field photography. PLOS one  \textbf{6}(12),  e27524 (2011)

\bibitem{graham2015kaggle}
Graham, B.: Kaggle diabetic retinopathy detection competition report. University of Warwick pp. 24--26 (2015)

\bibitem{hashir2020quantifying}
Hashir, M., Bertrand, H., Cohen, J.P.: Quantifying the value of lateral views in deep learning for chest x-rays. In: Medical Imaging with Deep Learning. pp. 288--303. PMLR (2020)

\bibitem{he2021multi}
He, X., Deng, Y., Fang, L., Peng, Q.: Multi-modal retinal image classification with modality-specific attention network. IEEE transactions on medical imaging  \textbf{40}(6),  1591--1602 (2021)

\bibitem{hou2022cross}
Hou, J., Xu, J., Xiao, F., Zhao, R.W., Zhang, Y., Zou, H., Lu, L., Xue, W., Feng, R.: Cross-field transformer for diabetic retinopathy grading on two-field fundus images. In: 2022 IEEE International Conference on Bioinformatics and Biomedicine (BIBM). pp. 985--990. IEEE Computer Society (2022)

\bibitem{li2021multi}
Li, X., Zhou, Y., Wang, J., Lin, H., Zhao, J., Ding, D., Yu, W., Chen, Y.: Multi-modal multi-instance learning for retinal disease recognition. In: Proceedings of the 29th ACM International Conference on Multimedia. pp. 2474--2482 (2021)

\bibitem{roh2021infrared}
Roh, H.C., Lee, C., Kang, S.W., Choi, K.J., Eun, J.S., Hwang, S.: Infrared reflectance image-guided laser photocoagulation of telangiectatic capillaries in persistent diabetic macular edema. Scientific reports  \textbf{11}(1),  21769 (2021)

\bibitem{rubin2018large}
Rubin, J., Sanghavi, D., Zhao, C., Lee, K., Qadir, A., Xu-Wilson, M.: Large scale automated reading of frontal and lateral chest x-rays using dual convolutional neural networks. arXiv preprint arXiv:1804.07839  (2018)

\bibitem{sayin2015ocular}
Sayin, N., Kara, N., Pekel, G.: Ocular complications of diabetes mellitus. World journal of diabetes  \textbf{6}(1), ~92 (2015)

\bibitem{sukkarieh2022role}
Sukkarieh, G., Lejoyeux, R., LeMer, Y., Bonnin, S., Tadayoni, R.: The role of near-infrared reflectance imaging in retinal disease: A systematic review. Survey of Ophthalmology  (2022)

\bibitem{wang2019two}
Wang, W., Xu, Z., Yu, W., Zhao, J., Yang, J., He, F., Yang, Z., Chen, D., Ding, D., Chen, Y., et~al.: Two-stream cnn with loose pair training for multi-modal amd categorization. In: Medical Image Computing and Computer Assisted Intervention--MICCAI 2019: 22nd International Conference, Shenzhen, China, October 13--17, 2019, Proceedings, Part I 22. pp. 156--164. Springer (2019)

\bibitem{10388423}
Wang, Y., Zhen, L., Tan, T.E., Fu, H., Feng, Y., Wang, Z., Xu, X., Goh, R.S.M., Ng, Y., Calhoun, C., Tan, G.S., Sun, J.K., Liu, Y., Ting, D.S.: Geometric correspondence-based multimodal learning for ophthalmic image analysis. IEEE Transactions on Medical Imaging pp.~1--1 (2024). \doi{10.1109/TMI.2024.3352602}

\bibitem{wilkinson2003proposed}
Wilkinson, C.P., Ferris~III, F.L., Klein, R.E., Lee, P.P., Agardh, C.D., Davis, M., Dills, D., Kampik, A., Pararajasegaram, R., Verdaguer, J.T., et~al.: Proposed international clinical diabetic retinopathy and diabetic macular edema disease severity scales. Ophthalmology  \textbf{110}(9),  1677--1682 (2003)

\bibitem{wu2019deep}
Wu, N., Phang, J., Park, J., Shen, Y., Huang, Z., Zorin, M., Jastrz{\k{e}}bski, S., F{\'e}vry, T., Katsnelson, J., Kim, E., et~al.: Deep neural networks improve radiologists’ performance in breast cancer screening. IEEE transactions on medical imaging  \textbf{39}(4),  1184--1194 (2019)

\bibitem{xue2022deep}
Xue, W., Zhang, J., Ma, Y., Hou, J., Xiao, F., Feng, R., Zhao, R., Zou, H.: Deep learning-based analysis of infrared fundus photography for automated diagnosis of diabetic retinopathy with cataracts. Journal of Cataract \& Refractive Surgery pp. 10--1097 (2022)

\bibitem{yu2021mil}
Yu, S., Ma, K., Bi, Q., Bian, C., Ning, M., He, N., Li, Y., Liu, H., Zheng, Y.: Mil-vt: Multiple instance learning enhanced vision transformer for fundus image classification. In: Medical Image Computing and Computer Assisted Intervention--MICCAI 2021: 24th International Conference, Strasbourg, France, September 27--October 1, 2021, Proceedings, Part VIII 24. pp. 45--54. Springer (2021)

\end{thebibliography}
\end{document}